\documentclass[12pt]{article}
\usepackage{amssymb}
\usepackage{amsfonts}
\usepackage{amsmath}
\usepackage{mathtools}  
\usepackage{bm}
\usepackage{latexsym}
\usepackage{epsfig}
\usepackage{bbm}
\usepackage{enumerate}
\usepackage{float}
\usepackage{color}
\usepackage{gensymb}
\usepackage{siunitx}
\usepackage{color}
\usepackage{hyperref}
\usepackage[round,authoryear]{natbib}
\usepackage[T1]{fontenc}
\usepackage[utf8]{inputenc}
\usepackage{authblk}
\usepackage{appendix}
\usepackage{blindtext}
\usepackage{lineno}
\usepackage{nth}
\usepackage{bbm}
\usepackage{setspace}

\newtheorem{definition}{Definition}

\usepackage[
        a4paper,
        left=2.5cm,
        right=2.5cm,
        top=3cm,
        bottom=3cm]{geometry}

\begin{document}

\title{Estimating Precipitation Extremes using the Log-Histospline}

\author{Whitney K. Huang\footnote{Statistical and Applied Mathematical Sciences Institute. E-mail: \href{whuang@samsi.info}{\nolinkurl{whuang@samsi.info}} and Pacific Climate Impacts Consortium, University of Victoria. E-mail: \href{whitneyhuang@uvic.ca}{\nolinkurl{whitneyhuang@uvic.ca}}}, Douglas W. Nychka\footnote{Department of Applied Mathematics and Statistics, Colorado School of Mines. E-mail: \href{nychka@mines.co}{\nolinkurl{nychka@mines.co}}}, Hao Zhang\footnote{Department of Statistics, Purdue University. E-mail: \href{zhanghao@purdue.edu}{\nolinkurl{zhanghao@purdue.edu}}}}

\date{\today}

\maketitle

\begin{abstract} 
One of the commonly used approaches to modeling extremes is the peaks-over-threshold (POT) method. The POT method models exceedances over a threshold that is sufficiently high so that the exceedance has approximately a generalized Pareto distribution (GPD). This method requires the selection of a threshold that might affect the estimates. Here we propose an alternative method, the Log-Histospline (LHSpline), to explore modeling the tail behavior and the remainder of the density in one step using the full range of the data. LHSpline applies a smoothing spline model to a finely binned histogram of the log transformed data to estimate its log density. By construction, a LHSpline estimation is constrained to have  polynomial tail behavior, a feature commonly observed in daily rainfall observations. We illustrate the LHSpline method by analyzing precipitation data collected in Houston, Texas.
\end{abstract}

%
%
%

\doublespacing
\section{Introduction} \label{sec1}
Estimating extreme quantiles is crucial for risk management in a variety of applications. For example, an engineer would seek to estimate the magnitude of the flood event which is exceeded once in 100 years on average, the so-called 100-year return level, based on a few decades of time series \citep{katz2002}. A financial analyst needs to provide estimates of the Value at Risk (VaR) for a given portfolio, essentially the high quantiles of financial loss \citep{embrechts1997,tsay2005}. In climate change studies, as the research focus shifts from estimating the global mean state of climate variables to the understanding of regional and local climate extremes, there is a pressing need for better estimation of the magnitudes of extremes and their potential changes in a changing climate \citep{zwiers1998,easterling2000,alexander2006,tebaldi2006,cooley2007,kharin2007,aghakouchak2012,shaby2012,huang2016}.\\

\begin{figure}[H]
\centering
\includegraphics[width=5in]{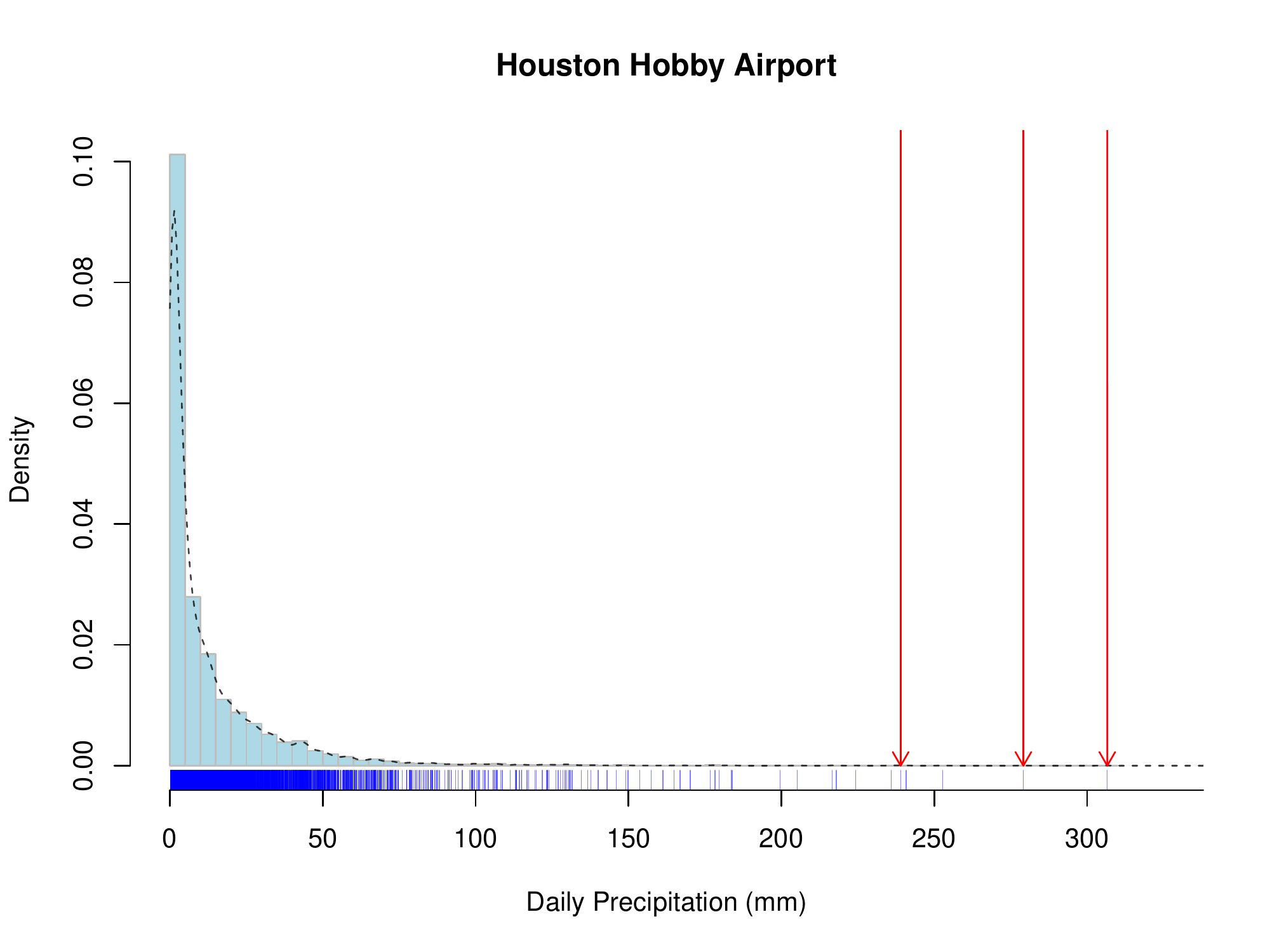}
\caption{Histogram of daily rainfall amount (mm) at the Hobby Airport in Houston, Texas, from 1949 to 2017. The vertical ticks at the x-axis are the values of the individual data points and the black dashed line is a kernel density estimate. Due to its tail heaviness, the largest values are substantially larger than the bulk of the distribution. Three out of the five largest precipitation measurements (indicated by red arrows) were observed during Hurricane Harvey, August 25 - 31, 2017}
\label{fig:fig1}
\end{figure}

The estimation of extreme quantiles poses a unique statistical challenge. Essentially, such an estimation pertains to the upper tail of a distribution where the available data of extreme values are usually sparse (e.g.\ see Fig.~\ref{fig:fig1}). As a result, the estimate will have a large variance that can increase rapidly as we move progressively to high quantiles. Furthermore, if the quantile being estimated is beyond the range of data (e.g., estimating the 100-year return level given the 50 years history of observation), the need to explicitly model the tail with some parametric form is unavoidable. \textit{Extreme value theory} (EVT) provides a mathematical framework of performing inference for the upper tail of distributions. One widely used approach, based on the extremal types theorem \citep{fisher1928,gnedenko1943}, is the so-called block maxima (BM) method where one fits a \textit{generalized extreme value} (GEV) distribution  to block maxima given that the block size is large enough. The reader is referred to \citep{jenkinson1955}, \cite{gumbel1958} and \cite{coles2001} for more details.

One drawback of the BM method is that data are not used efficiently, that is, the block size $b$ needs to be sufficiently large so that the GEV distribution is approximately valid for a given time series (with sample size $n$), which makes the sample size of the block maxima $(n/b)$ substantially smaller than the original sample size. Moreover, the top few largest order statistics within a given block should, in principle, inform us about the behavior of extreme events \citep{weissman1978,smith1986}. The peak-over-threshold (POT) method is based on the Pickands--Balkema--de Haan theory \citep{pickands1975, balkema1974}. For a random variable $Y$ with cumulative distribution function $F(\cdot)$, it states that the distribution of the exceedance over the threshold (i.e., the distribution of $Y-u$ given $Y>u$) converges to a \textit{generalized Pareto distribution} (GPD) as the threshold $u$ tends to the upper limit point $y_{F} = \sup\{ y: F(y) < 1\}$. With an appropriately chosen $u$, the POT method makes use of the exceedances and the censored data of those that do not exceed the threshold \citep{smith1984, davison1984,davison1990}. One advantage of the POT method over the BM method is that it typically makes use of the available data more efficiently in estimating extreme events.\\

However, to apply the POT method, a threshold $u$ has to be chosen and the estimates may be sensitive to the chosen threshold \citep{scarrott2012, wadsworth2012a}. The threshold selection involves a bias-variance trade-off: if the chosen threshold is too high, the estimates exhibit large variability; if the chosen threshold is too low, the asymptotic justification of the GPD approximation to the tail density is less effective, which leads to bias. There are several graphical tools that aim to help with the threshold selection, for example, the mean residual life plot and the parameter stability plot \citep[e.g.][p.80 and p.85]{coles2001}. However, the use of these graphical tools does not always lead to a clear choice of the threshold. In general, automated threshold selection is a difficult problem \citep{dupuis1999,wadsworth2012a}. In addition, the POT method does not assume anything about the distribution below the chosen threshold, which can be of interest in some applications (e.g., stochastic weather generators \citep{kleiber2012,li2012}), although some efforts have been made. We will review some attempts in the next paragraph.\\

Recently, there are some attempts to model the distribution of a random variable while retaining a GPD tail behavior \citep{frigessi2002,tancredi2006,carreau2009,papastathopoulos2013,naveau2016}. These methods can be broadly divided into two categories: the extreme value mixture models \citetext{see \citealp{scarrott2012,dey2015} for review} and the \textit{extended generalized Pareto distribution} (EGPD) method \citep{papastathopoulos2013,naveau2016}. The basic idea of the extreme value mixture is to model a distribution as a mixture of a \textit{bulk} distribution and a GPD distribution above a threshold. The threshold is then treated as an additional parameter to be estimated from the data. However, the specification of the bulk distribution can have non-negligible impacts on the estimates of GPD parameters, which can lead to substantial biases in the tail estimates. Finally, the estimation uncertainty for the threshold can be quite large \citep{frigessi2002} and hence it is not clear whether this parameter can be identifiable. The extended Pareto method proposed by \cite{papastathopoulos2013} bypasses this issue of the mixture modeling by proposing several classes of parametric models with GPD limiting behavior for the upper tail. \cite{naveau2016} extended the scope of this approach by allowing the classes of parametric models with GPD limiting behavior for both lower and upper tails in an application to rainfall modeling.\\

This work presents a new statistical method, called the Log-Histospline (LHSpline), for estimating probabilities associated with extreme values. Similar to \cite{naveau2016}, this method applies  extreme value theory to the upper tail distribution. However, it does not impose a parametric form in the bulk of the distribution where the density can be determined from the data. This work is motivated by some applications in climate studies, one of which involves precipitation data. We would like to (i) provide a flexible model to the \textit{full range} of the non-zero rainfall distribution, and (ii) reliably estimate extreme events (e.g.\ 100-year daily rainfall amount).  Specifically, our approach is to first log transform the nonzero daily rainfall observations and then apply a generalized cubic smoothing spline on a finely binned histogram to estimate the log-density. The purpose of the data transformation step is similar to that of a variable bandwidth in  density--estimation literature \citep{wand1991}. Applying the spline smoothing log-density estimation \citep{silverman1982,gu1993} to the transformed variable will ensure the algebraic (e.g., Pareto) upper tail behavior and gamma-like lower (near 0) tail behavior, commonly observed in daily rainfall data.\\

The rest of this paper is structured as follows: In Section~\ref{sec2}, the LHSpline method is introduced and computation for inferences is described. A simulation study is presented in Section~\ref{sec3}. An application to daily precipitation collected in Houston Hobby Airport, Texas is illustrated in Section~\ref{sec4}. Some discussion of future research directions is provided in the last section.

\section{The Log-Histospline} \label{sec2}

\subsection{The Model of Log Density: Natural Cubic Spline}
Let $Y$ be a continuous random variable with a probability density function (pdf) $f(y)$. Throughout this work we assume that $Y \in (0, \infty]$ is heavy-tailed, i.e.,
\begin{equation}
f(y) \sim y^{-(\alpha+1)} \text{ as } y \to \infty, \mbox{ for some } \alpha >0.     
\end{equation}  

To represent the heavy tail of the distribution of $Y$,  we take the logarithmic transformation $X = \log (Y)$, via its log-density, and assume the pdf of $X$ takes the following form:
\begin{equation}
e^{g(x)}, \quad x \in (-\infty, \infty]
\end{equation}  
Specifically, 
the density $f(\cdot)$ and log-density $g(\cdot)$ are related in the following way:
\begin{equation} \label{pdf-transform}
f(y) =  y^{-1}e^{g(\log(y))}, \qquad  y > 0.
\end{equation} 

Modeling the log-density $g$ has an advantage that it conveniently enforces the positivity constraint on $f$
\citep{leonard1978,silverman1982,kooperberg1991,eilers1996}. Here we would like to model $g$ nonparametrically to avoid a strong and likely misspecified parametric structure \citep[e.g.][]{silverman1986}. However, it is well-known that the usual nonparametric density estimation procedure will often introduce spurious features in the tail density due to limited data in the tail \citep[e.g.][]{kooperberg1991}. This issue is amplified when dealing with heavy-tailed distributions that are typically observed in precipitation data. On the other hand, since the focus is on estimating extreme quantiles, it is critical that the model can capture both qualitative (polynomial decay) and quantitative (the tail index $\alpha$) behaviors of the tail density. Here we assume that $g$ belongs to the family of \textit{natural cubic splines} (see Definition~\ref{ncp}) to accommodate both a flexible bulk distribution and the algebraic tail. The LHSpline combines the ideas of \textit{spline smoothing} \citep[e.g.][]{wahba1990,gu2013} and \textit{Histospline} \citep{boneva1971} to derive an estimator to achieve these goals.
To facilitate the description of our method, we review the following definition \citep{wahba1990}
\begin{definition}[\textbf{Natural cubic spline}] \label{ncp}
A natural cubic spline $g(x)$, $x \in [a,b]$ is defined with a set of points $\{\zeta_{i}\}_{i=1}^{N}$ (knots) such that $-\infty \leq a \le \zeta_{1} \le \zeta_{2} \le \cdots \le \zeta_{N} < b \leq \infty$ with the following properties:
\begin{enumerate}
\item $g$ is linear, $\quad x \in [a, \zeta_{1}], \quad x \in [\zeta_{N}, b]$
\item $g$ is a cubic polynomial, $\quad  x \in [\zeta_{i}, \zeta_{i+1}],  \quad i = 1, \cdots, N -1$
\item $g \in \mathcal{C}^{2},  \quad x\in (a, b)$
\end{enumerate}   
where  $\mathcal{C}^{k}$ is the class of functions with $k$ continuous derivatives 
\end{definition}

The cubic spline smoother is widely used in nonparametric density estimation due to its flexibility  and hence provides a flexible model for the bulk of the distribution \citep[e.g.][]{silverman1986,gu2013}. 
In our setting,  when $y \ge \exp(\zeta_{N})$, the tail density of $Y$ is  
\begin{align*}
f(y) & = y^{-1}e^{g(\log(y))} = y^{-1}e^{-\alpha \log(y) + \beta}\\
& = y^{-1} e^{-\alpha \log(y)}e^{\beta} = Cy^{-1}e^{-\alpha \log(y)}\\
& = Cy^{-1}y^{-\alpha} = Cy^{-(\alpha + 1)} \tag{4}
\end{align*}
where $-\alpha$ is the slope and $\beta$ is the intercept of the log-density $g$ at $\zeta_{N}$. Therefore, by the linear boundary conditions on a cubic spline we automatically obtain the algebraic right tail behavior. 

\subsection{The Estimation Procedure} \label{sec:2.2}

\begin{figure}[H]
\centering
\includegraphics[width=5.5in]{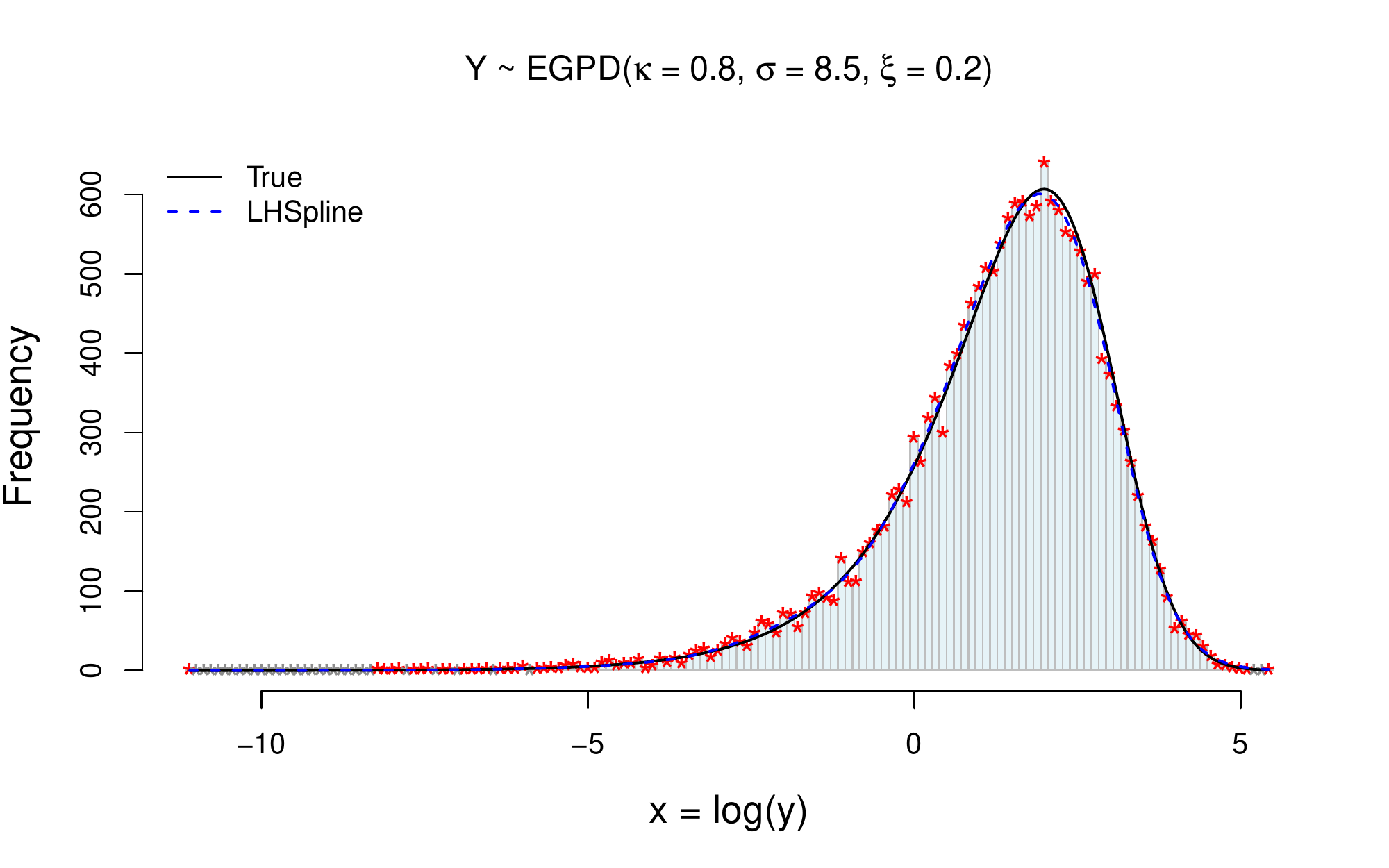}
\caption{An illustration of applying LHSpline to a simulated data set. Here we bin the log transformed data $\{x_{i}\}_{i=1}^{18,250}$ into $N = 150$ equally-spaced intervals to obtain a histogram. Red asterisks denote the non-zero counts and gray asterisks denote the zero counts. The LHSpline estimate (blue dashed line) is obtained by fitting a penalized Poisson regression to the histogram counts (both red and gray asterisks); the true density (the density of $X = \log(Y)$ where $Y \stackrel{i.i.d}{\sim} \text{EGPD}(\kappa = 0.8, \sigma = 8.5, \xi = 0.2)$) is plotted as black solid line.}
\label{fig:fig2}
\end{figure}

Given an i.i.d.\ sample $Y_{1}, Y_{2}, \ldots, Y_{n}$ of $Y$, the estimation of $g$ involves two steps: data binning to create a histogram object, and a generalized smoothing spline of the histogram. In what follows, it is useful (see Fig.~\ref{fig:fig2}) to illustrate these steps using a synthetic data set simulated from a EGPD \citetext{\citealp{naveau2016}, p.\ 2757}. Some considerations when applying LHSpline to precipitation data will be discussed in Section~\ref{sec4}.\\

\textbf{Data binning}: First, bin the transformed data $X_i=\log(Y_i), i=1, \ldots, n$ by choosing $N+1$ equally spaced break points  $\{b_j \}$ to construct the corresponding histogram object (i.e.\ the histogram counts $\{Z_{j}\} = \# \{ X_{i}: X_{i}\in [b_{j}, b_{j+1}]\}$. 
We set the knots associated with the spline to the bin midpoints:
$\zeta_j = (b_{j} + b_{j+1})/2$. Several remarks on data binning should be made here:
\begin{enumerate}[(1)]
\item Equal-sized bins in the log scale implies a variable bin size with the bin size becoming increasingly larger in the upper tail.
\item The number of bins should be ``sufficiently'' large for the Poisson assumption made in the next step justifiable.
\item The choice of the first and especially the last break points (i.e.\ $b_{1}$ and $b_{N+1}$) will have non-negligible impacts on tail estimation in our framework. We will choose them somehow smaller (larger) than the minimum $X_{(1)}$ (maximum $X_{(n)}$) of the log-transformed data, which is different than what is typically done in constructing a histogram. We will defer the discussion on this point to Sec~\ref{sec2.3}.   
\item It is assumed  that the bin size is fine enough so that  $\int_{b_{i}}^{b_{i+1}} e^{g(x)} dx $ is well approximated by 
$e^{g(\zeta_i)}(b_{i+1} - b_{i}).$
\end{enumerate}

\textbf{Smoothing the histogram}: We adapt a penalized approach \citep{good1971,tapia1978,green1994} to obtain a functional estimate of $g$ as follows. First, we assume the bin counts $\{Z_{j}\}_{j=1}^{N}$ each follow a Poisson distribution\footnote{\{$Z_{j}\}_{j=1}^{N}$ has a \textit{multinomial} distribution with parameters $n$ and $\bm{\pi} = (\pi_{1}, \cdots, \pi_{N})$ where $\pi_{j}$ is the probability that a random variable $X$ falls into the j$_{th}$ bin. Here we consider $Z_{j} \stackrel{ind}{\sim}\text{Pois}(\gamma \pi_{j}), \, j = 1,\cdots, N$. The MLE of $\gamma$ under the Poisson model follows $\hat{\gamma} = \sum_{j=1}^{N}Z_{j} = n$ and $\hat{\bm{\pi}}$ is equal to the MLE of $\bm{\pi}$ under the multinomial model.} \citep{lindsey1974a,lindsey1974b,eilers1996,efron1996} with log intensity $\tilde{g}_{j}, \, j = 1, \cdots, N$, given that $N$ is large enough so that the bin size is fine enough. Furthermore, we assume $\tilde{g}_{j} \approx \tilde{g}(\zeta_j)$, i.e., we assume the Poisson log intensity at each bin represents the log intensity function (unnormalized log density function) evaluated at its midpoint. Hence we perform a penalized Poisson regression to the data pair $\{Z_{j}, \zeta_j\}_{j=1}^{N}$ using a log link function and with a penalty term that penalizes the ``roughness'' of $\tilde{g}$. The estimate $\tilde{g}$, the unnormalized version of $g$, can be obtained by solving the following minimization problem:

\begin{equation}\label{pnl}
-\sum_{j=1}^{N}
\left\{
\tilde{g}(x_j)z_{j}-e^{\tilde{g}(x_j)} -\log(z_{j})!
\right\} 
+ \lambda\left(\int _{x \in \mathbb{R}} \left(\tilde{g}^{\prime \prime}(x)\right)^2\,dx\right)
\end{equation}
where $\lambda$ is the smoothing parameter. Note that the first term in the objective function is the negative log likelihood for the histogram counts under a Poisson approximation to the distribution, and the second term, the squared integral of the second derivative of $\tilde{g}$ multiplied by the smoothing parameter, is the penalty function. The solution to this optimization problem exists, is unique, and has the form of a natural cubic spline for any $\lambda$ \citep{green1994,gu2013}. The selection of the smoothing parameter $\lambda$ plays a role of balancing the data fidelity of the Poisson regression fit to the histogram counts, as presented by the negative log-likelihood, and the ``smoothness'' of $\tilde{g}$, and is chosen by approximate cross validation \citetext{CV, see \citealp{o1988} for more details}. Note that the smoothness penalty favors linear functions, that correspond to the algebraic tail behavior we want in the untransformed distribution. Finally we renormalize the $\tilde{g}$ to make the integral of $\int_{\mathbb{R}} e^{g}$ equal to one. Numerical integration is used to approximate the integral within the data range and an analytic form is used for the density beyond the bin limits.

\subsection{Bias Correction} \label{sec2.3}

\subsubsection{Extending Data Range} \label{sec2.3.1}

As mentioned in the remark (3) in the data binning step, the choice of $b_{1}$ and $b_{N+1}$ plays an important role in our penalized approach. Under the usual histogram construction (i.e.\ $b_{1} = X_{(1)}$, $b_{N+1} = X_{(n)}$, see Fig~\ref{fig:fig2}) and a large number of bins, as in our setting, one will very likely observe ``bumps'' (i.e., isolated non-zero counts) in the boundaries of the histogram. As a result, the cross validation will maintain the smoothness and tend to overestimate the slopes near the boundary knots. Our solution to reduce this bias is to extend the range of the histogram beyond the sample maximum/minimum so that some zero counts beyond the boundaries will be included. In fact, one can think of the estimation procedure of the LHSpline as a discrete approximation to estimating the intensity (density) function of an  (normalized) inhomogeneous Poisson process \citep{brown2004}. In this regard, one should take into account the support (observation window in the point process context) as part of the data. 

\subsubsection{Further Corrections: Bootstrap and Smoothing Parameter Adjustment} \label{sec2.3.2}

After applying the aforementioned boundary correction by extending the observation window, tail bias still exists.  Here we propose two approaches to correct the bias. The first approach is to estimate the bias via a ``parametric''  bootstrap \citep{efron1979} on the LHSpline estimate. Note that given the approximate Poisson model for the histogram bin counts it is straightforward to generate bootstrap samples for this computation. The bias is estimated as the difference between the (point-wise) mean of the bootstrap estimates $\bar{g^{*}}$ and the ``true'' $g$, in this case, $\hat{g}$ estimated from the histogram.  
We then subtract this bias term from the LHSpline estimate.  The second approach is to adjust the smoothing parameter estimate $\hat{\lambda}_{\text{CV}}$ directly. We found that the smoothing parameter obtained from CV may not give the best result as here the primary objective is to estimate the tail density. It was found that, in our LHSpline, reducing $\lambda$,  \ $0.05 \times \hat{\lambda}_{\text{CV}}$,  gives a better tail estimation while still maintaining a good estimation performance in the bulk distribution (see Sec~\ref{sec3.3}).    

\subsection{Quantifying Estimation Uncertainty: The ``Bayesian'' Approach} \label{sec2.4}

Here we give a brief account of the Bayesian interpretation of smoothing splines that will be used for quantifying estimation uncertainty for the LHSpline. Readers are referred to \citetext{\citealp{wahba1990}, Chapter 5, \citealp{green1994}, Chapter 3.8, and \citealp{gu2013}, Chapter 2.4} for more details.\\

From a Bayesian perspective, the penalty term in Equation~(\ref{pnl}) effectively introduces a Gaussian process prior on $g$, the log density. Hence the LHSpline estimation can be thought as a procedure of finding the posterior mode of $g$ with a Poisson likelihood for the histogram and a zero mean Gaussian process with a generalized covariance (the reproducing kernel), $K(\cdot,\cdot)$, as the prior. The LHSpline estimate of a given histogram can be found by using the Fisher scoring algorithm \citetext{\citealp{green1994}, p.\ 100} where the minimization of the non-quadratic negative penalized log-likelihood is approximated by the method of iteratively reweighted least squares (IRLS).\\

The goal here is to sample from an approximate posterior distribution $[\bm{g}|\bm{u}, \lambda]$ where $\bm{g} = \{g _{k} = g(\zeta_k)\}_{k=1}^{N}$, $\bm{u} = \{u_{k}\}_{k=1}^{N}$ are the ``pseudo'' observations in IRLS, $\lambda$ is the smoothing parameter, and the likelihood has been approximated with a multivariate Gaussian distribution. Based on the linear approximation around the posterior model we have
\begin{equation}
[\bm{g}|\bm{u}, \lambda] \sim \text{MVN}(\hat{\bm{g}}, (W + \Gamma)^{-1})
\end{equation}
where $W$ is a diagonal matrix representing the approximate precision matrix of the pseudo observations and $\Gamma$ is the prior precision for $g$ based on the Bayesian interpretation of a spline. Several assumptions are being made here: 1) the smoothing parameter $\lambda$ is known; 2) the linear problem is assumed to be multivariate normal and finally 3) a proper prior, $\text{N}(\bm{0}, \epsilon I)$, for the linear trend (i.e., the null space of a natural cubic spline) has been taken in the limit to be improper (i.e., $\epsilon \to \infty$).\\

The sampling approach is similar to a classical bootstrap \citep{efron1979} with the following steps: 
\begin{enumerate}
\item generate $\bm{g}^{*} \sim \text{MVN}(\bm{0}, \Gamma^{-1})$
\item generate pseudo observations $\bm{u}^{*} \sim \text{MVN}(\bm{g}^{*}, W^{-1})$
\item compute the estimate based on $\bm{u}^{*}$: $\hat{\bm{g}}^{*} = (W+\Gamma)^{-1}W\bm{u^{*}}$
\item compute error: $\bm{u} = \bm{g}^{*} - \hat{\bm{g}}^{*}$
\item approximate draw from the posterior, $\hat{\bm{g}} + \bm{u}$, where $\hat{\bm{g}}$ is the LHSpline estimate (the posterior mode/mean under the aforementioned Gaussian assumption) evaluated at knots $\{\zeta_{k}\}_{k=1}^{N}$. 
\end{enumerate}

\section{Simulation Study} \label{sec3}

The purposes of this simulation study are threefold: (i) to demonstrate how we implement the LHSpline, (ii) to compare the LHSpline method with the EVT-based methods (i.e.\ POT and BM) in estimating extreme quantiles, and (iii) to compare with a gamma distribution fit to the bulk distribution.

\subsection{Setup} \label{sec3.1}
We conduct a Monte Carlo study by simulating the data from the model (i) of the EGPD \citetext{see \citealp{naveau2016}, p.\ 2757}. The basic idea of EGPD is to modify the GPD random number generator $H_{\sigma, \xi}^{-1}(U)$, where $H^{-1}_{\sigma, \xi}$ denotes the inverse GPD cdf with scale $(\sigma)$ and shape $(\xi)$ parameters, by replacing $U$, the standard uniform random variable, with $V = G^{-1}(U)$ where $G$ is a continuous cdf on $[0,1]$. The model (i) takes $G(v) = v^{\kappa}, \kappa>0$ which contains the GPD as a special case (i.e., $\kappa = 1$).

We choose the parameter values to be $\kappa = 0.8, \sigma = 8.5, \xi =0.2$ and the sample size $n = 18,250$, which corresponds to a time series with 50 years of complete daily data (ignore the leap years). The EGPD parameters are chosen to reflect typical distributional behavior of daily precipitation. We repeat the Monte Carlo experiment 100 times and evaluate the estimation performance for $25-$, $50-$, and $100-$ year return levels.

\subsection{LHSpline Illustration} \label{sec3.2}
\begin{figure}[H]
\centering
\includegraphics[width=6in]{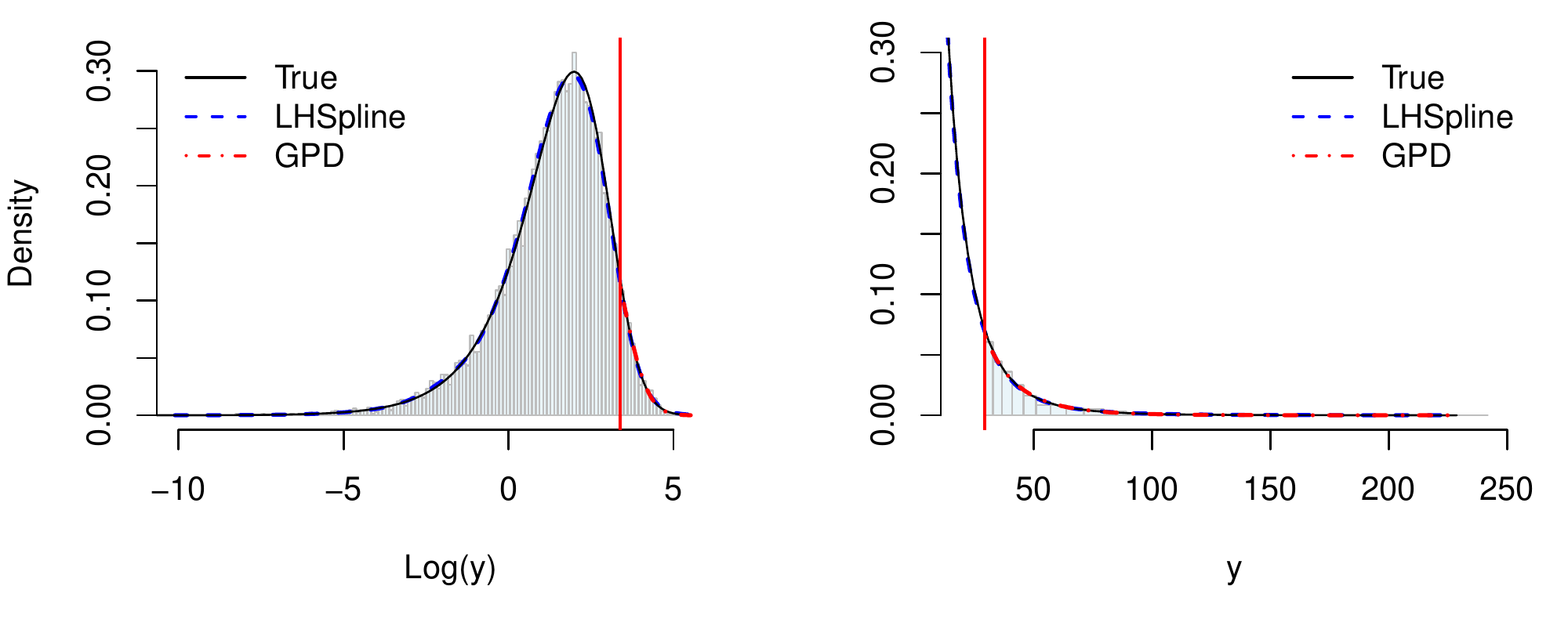}
\caption{The density estimates for log scale $X$ \textbf{(left)} and original scale $Y$ \textbf{(right)}. In each panel, the blue dashed line is the LHSpline estimate (without boundary correction), red dot-dash POT estimate with $u$ as the .95 empirical quantile, and black truth density function.}
\label{fig:fig3}
\end{figure}

We use a simulated data series (see Fig.~\ref{fig:fig2}) to illustrate the LHSpline method as described in Sec.~\ref{sec2}. To illustrate the ``boundary effect'' we first set equidistant breakpoints so that the knots (the midpoints of the histogram) span the whole range of the data (i.e.\ $\zeta_{1} = x_{(1)}$, $\zeta_{N}=x_{(n)}$ where $\zeta_{1}$ and $\zeta_{N}$ are the first and last knot points). We choose $N = 150$ meaning that we construct a histogram with a rather large number of bins (much larger than what the default in \texttt{hist} in \textsf{R} would suggest). We then apply the generalized smoothing spline procedure by solving equation~(\ref{pnl}) via approximate cross-validation to obtain an estimate. Figure ~\ref{fig:fig3} shows the LHSpline estimate along with a GPD fit (with threshold $u$ chosen as the $.95$ empirical quantile) and the true density. Upon visual inspection one may conclude that LHSpline performs well, or at least as good as the POT method, and both estimates are fairly close to the true density. 

\begin{figure}[H]
\centering
\includegraphics[width=6in]{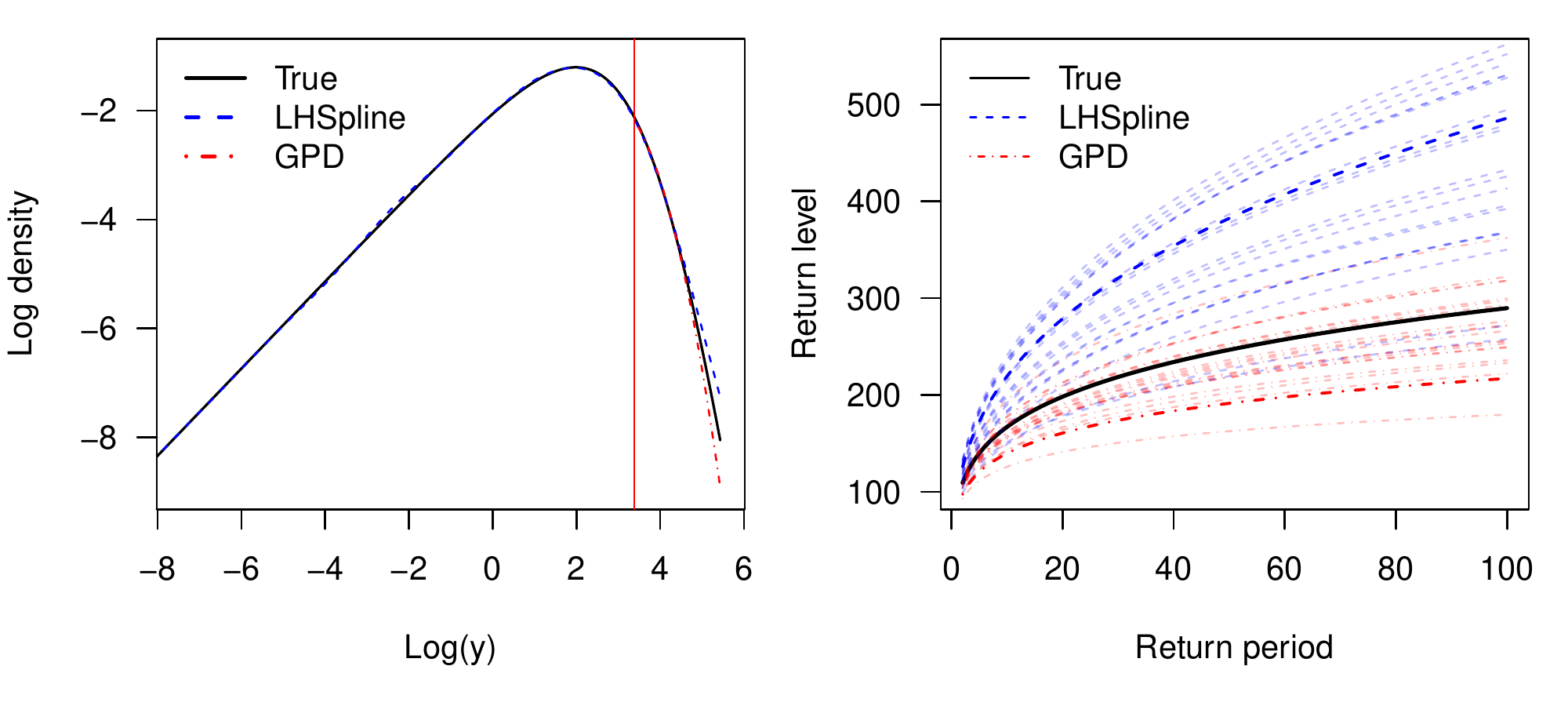}
\caption{\textbf{Left}: Estimated log density against $\log (y)$. \textbf{Right}: Estimated return levels with return period ranging from 2 years to 100 years. The light blue dashed (red dot-dash) curves are 20 samples from the 100 Monte Carlo experiments for GPD fits with threshold $.95$ quantile and LHSpline fits without boundary correction.}
\label{fig:fig4}
\end{figure}

However, a more careful examination for the log-log plot (log-density against $\log(y)$, see the left panel of Fig~\ref{fig:fig4}) and the return levels estimation reveals that LHSpline with $\zeta_{1} = x_{(1)}$ and $\zeta_{N}=x_{(n)}$, in general, overestimates the extreme quantiles (see Fig.~\ref{fig:fig4}, right panel).

The issue of overestimation of LHSpline is partly due to the ``boundary effect'' when fitting penalized Poisson regression to a histogram. Specifically, the histogram counts in Fig.~\ref{fig:fig2} are $\{Z_{1} = 1, Z_{2} = Z_{3} = \cdots = Z_{26} = 0, \cdots,Z_{148} = Z_{149} = 0, Z_{150} = 1\}$. The histogram counts in the first and the last bins are non-zero (typically 1) by construction but the nearby bins are likely to be zero. Therefore, these ``bumps'' at both ends force the smoothing spline to overestimate the slopes at the boundaries (sometimes it will give nonsensical results, e.g., negative slope at the left boundary or/and positive slope at the right boundary) and hence the extreme quantiles.\\ 

\begin{figure}[H]
\centering
\includegraphics[width=5.5in]{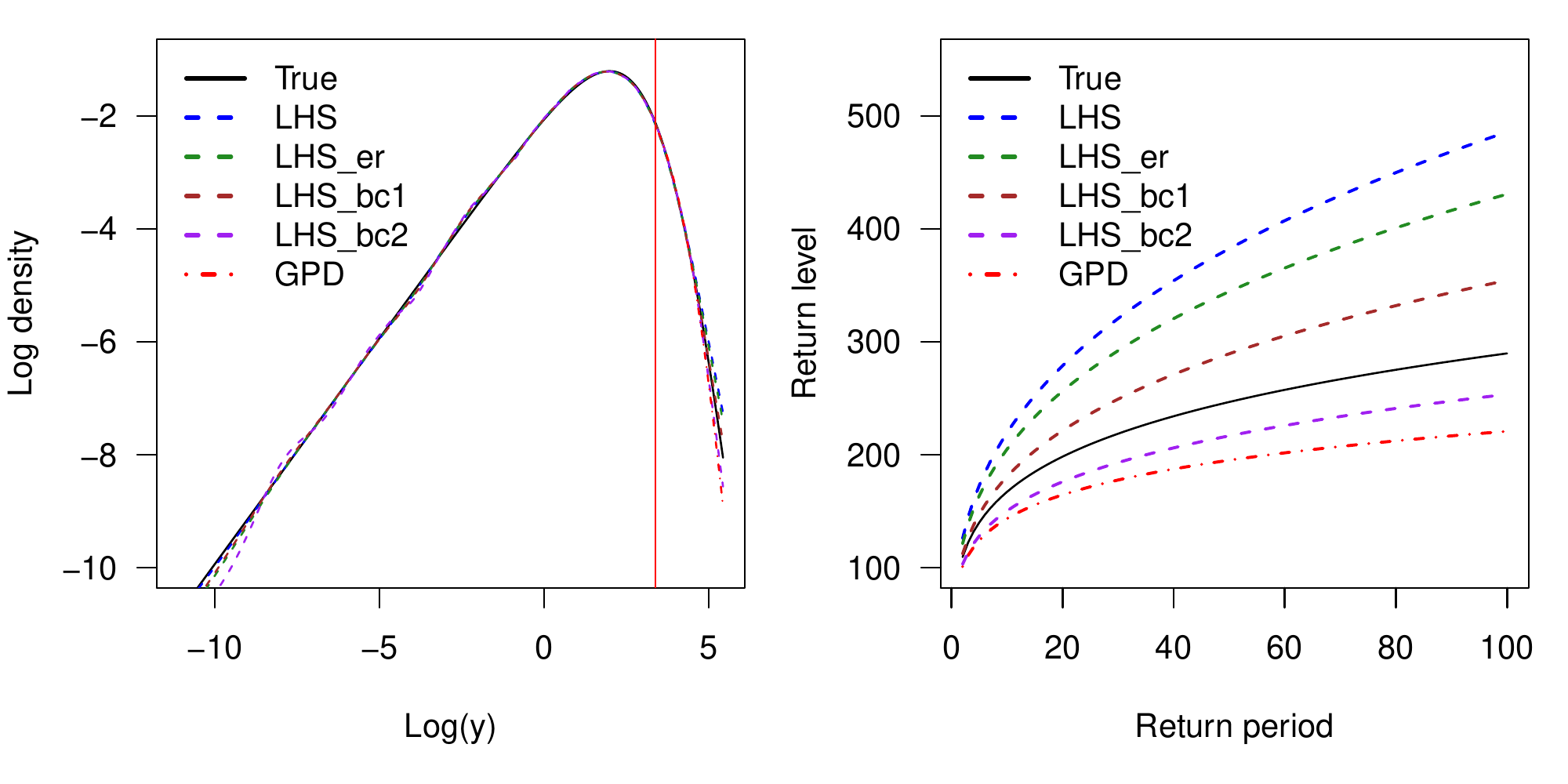}
\caption{The estimated log density and return levels with (green dashed line, LHS-er) and without (blue dashed line, LHS) boundary correction. Brown (purple) dashed lines is the LHSpline estimate with  bootstrap ($\lambda$ adjustment) bias correction. The GPD estimate is shown in the red dot-dash line. The vertical red line is the chosen threshold for fitting the GPD.}
\label{fig:fig5}
\end{figure}

To alleviate this boundary effect, we extend the range and number of bins in the histogram beyond the range of the data to include extra bins that have zero counts and refit these augmented data pairs (the original counts and those extra zero counts) to penalized Poisson regression. We observe that this strategy does remove some of the positive bias in return levels estimation (see Fig.~\ref{fig:fig5}) but there is still some positive bias remaining. We then apply the bias correction via bootstrap with 200 bootstrap samples and the aforementioned smoothing parameter adjustment. Both approaches further reduce the bias (see Fig~\ref{fig:fig5}). 

\subsection{Estimator Performance} \label{sec3.3}

In this subsection we first assess the performance of estimating high quantiles using an LHSpline and two commonly used EVT-based methods, namely, the block maxima (BM) method by fitting a GEV to ``annual maxima'', and the peaks-over-threshold (POT) method by fitting a GPD to excesses over a high threshold ($.95$ empirical quantile in our study). In order to put this comparison on an equal footing as much as possible, we put a positive Gamma prior (with mean equal to 0.2, the true value) on $\xi$
when fitting the GEV and GPD models.

We use the \texttt{fevd} function in the \texttt{extRemes}
R package \citep{extRemes} with the  Generalized maximum likelihood estimation (GMLE) method \citep{martins2000,martins2001} to estimate the GEV and GPD parameters.
We compare the GMLE plug-in estimates for 25-, 50-, and 100-year return levels with that of the LHSpline estimates (with and without bias correction). To get a better sense of how well each method performs, we also include the ``oracle estimates'' where we fit the true model (EGPD) using the maximum likelihood method to obtain the corresponding plug-in estimates.\\ 

\begin{figure}[H] 
\centering
\includegraphics[width=6in]{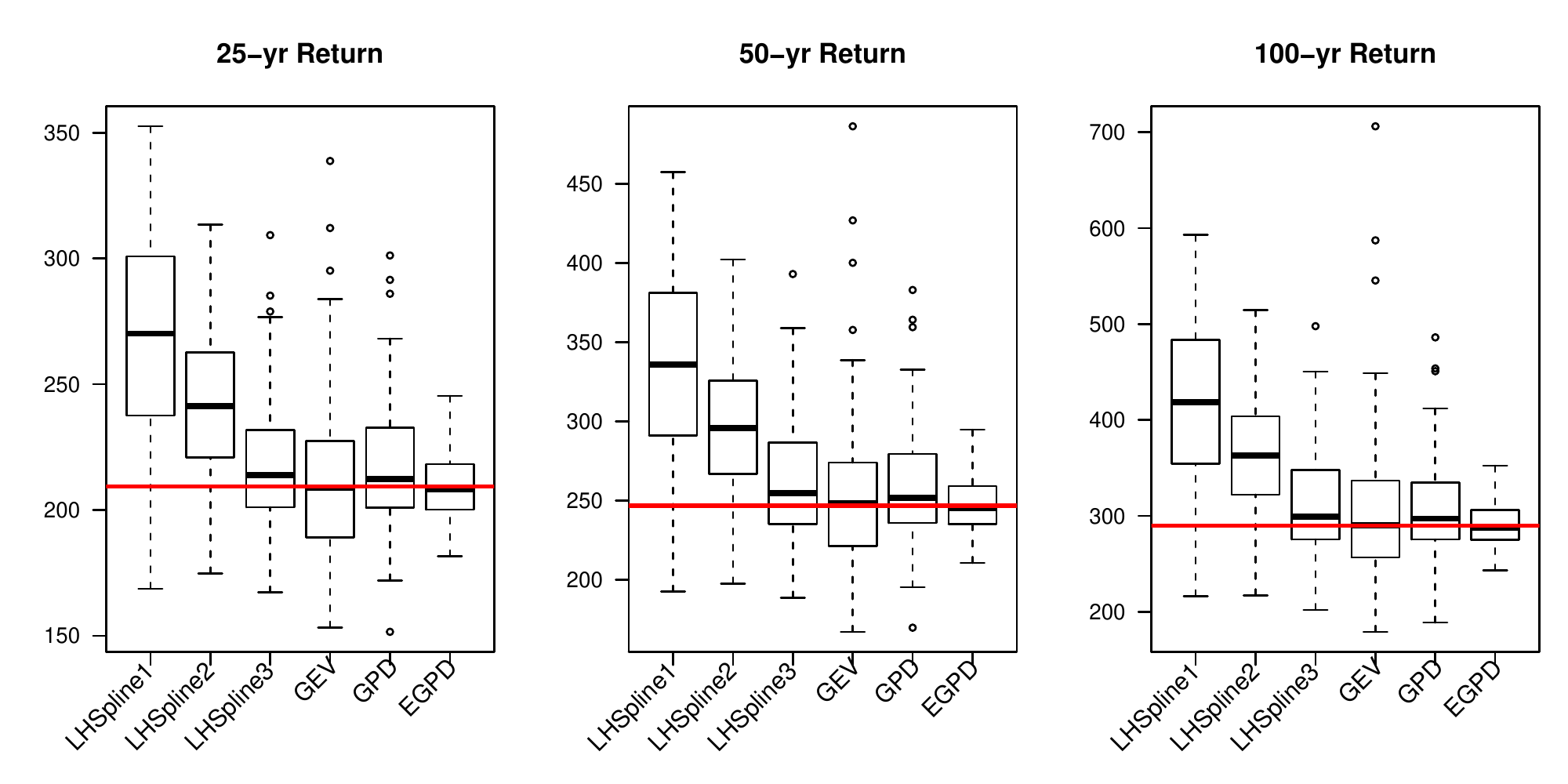}
\caption{Boxplots of estimated \textbf{Left:} 25-, \textbf{Middle:} 50-, and \textbf{Right:} 100-year return levels using (from left to right in each panel) \textbf{LHSpline1:} LHSpline with boundary correction (extend data range by a factor of 1.5), \textbf{LHSpline2:} LHSpline with boundary correction and bootstrap bias correction, \textbf{LHSpline3:} LHSpline with boundary correction and smoothing parameter adjustment $(\lambda = 0.05 \times \hat{\lambda}_{\text{CV}})$, \textbf{GEV:} block maxima method, \textbf{GPD:} Peaks-over-threshold method, and \textbf{EGPD} the oracle (extended GP). Boxplots are based on 100 independent replicates, and true values are represented by horizontal red lines.}
\label{fig:fig6}
\end{figure}

Figure~\ref{fig:fig6} shows that, perhaps not surprisingly, the variability of the return level estimates obtained by fitting a GEV distribution to block maxima is generally larger than that of the estimates obtained by fitting a GPD distribution to threshold exceedances under the i.i.d.\ setting here. Also, it is clear that the LHSpline without bias correction can not only lead to serious bias but also inflate the estimation variance. After applying two different bias corrections mentioned in Sec~\ref{sec2.3}, the estimate becomes closer to being unbiased (especially the bias correction with adjustment of the smoothing parameter) with a reduction in the estimator variability. Also the estimator performance is comparable with that of the GPD.

\begin{figure}[H] 
\centering
\includegraphics[width=3in]{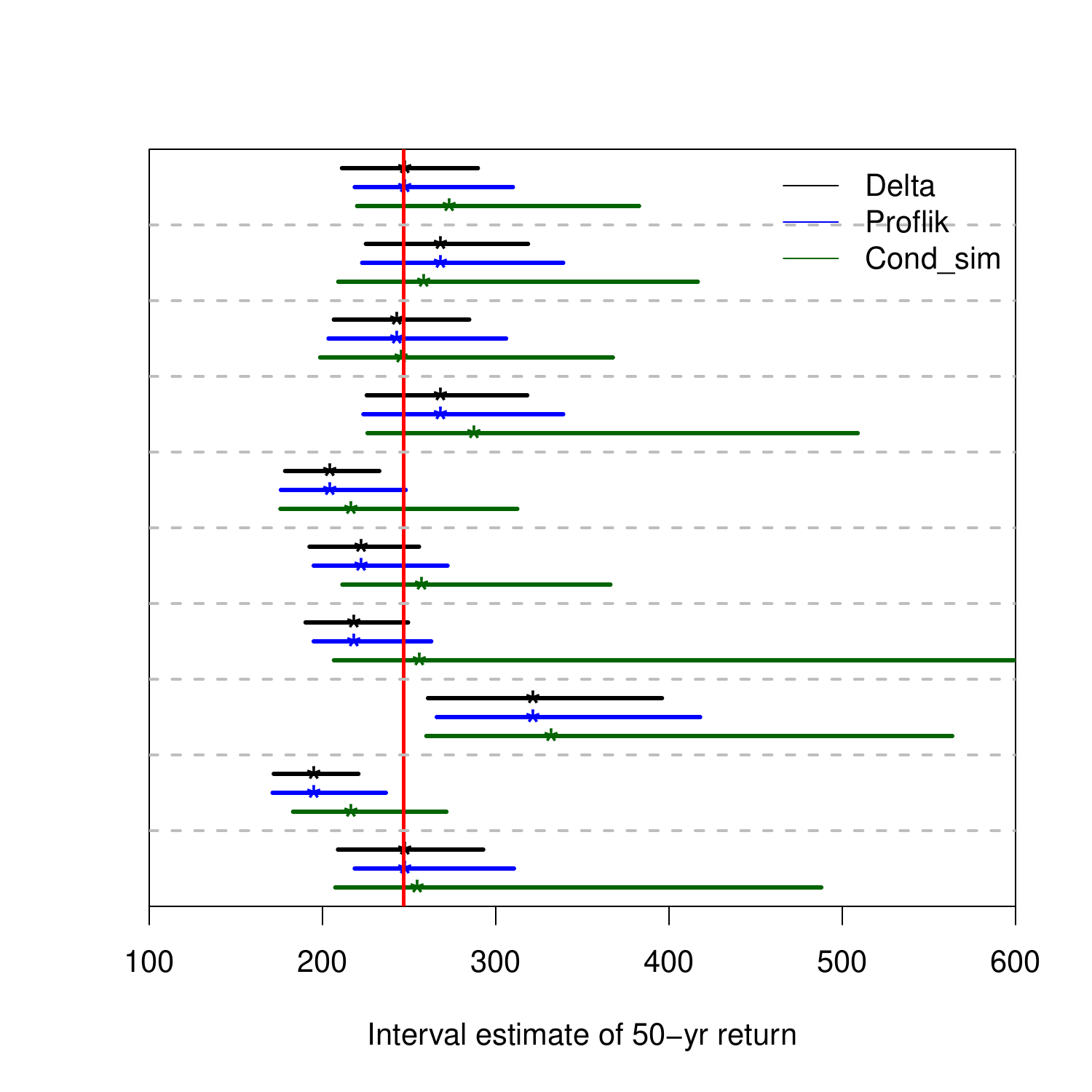}
\caption{The interval estimate of LHSpline (with $\lambda = 0.05 \times \hat{\lambda}_{CV}$) and GPD (with $u=$ .95 quantile). Each black (blue) horizontal line is the 90 \% CI using the delta (profile likelihood) method for simulated data. The darkgreen horizontal lines are the 90 \% CI of the smoothing parameter  adjusted LHSpline fit using the conditional simulation approach described in Sec~\ref{sec2.4}. The vertical red line denotes the true 50-year return level. The gray horizontal dashed lines are used to divide 10 different Monte Carlo draws.}
\label{fig:fig7}
\end{figure}

We then assess the estimation uncertainty using the conditional simulation approach described in Sec~\ref{sec2.4} and compare with the GPD interval estimation obtained by the delta and profile likelihood methods. To simplify this presentation, we only show the result of the 90\% confidence interval for the 50-year return level (see Fig~\ref{fig:fig7})

We summarize in Table~\ref{table1} the empirical coverage probability (ECP) and the width of CI for the  delta and profile likelihood methods when fitting a GPD to excesses over the .95 quantile and the conditional simulation (cond-sim) when fitting LHSpline with $\lambda = 0.05 \times \hat{\lambda}_{\text{CV}}$ for 25-, 50-, and 100-year return levels, respectively.

\begin{table}[H]
\caption{The empirical coverage probability (ECP) and (90\%) CI width in parentheses for each method for 25-, 50-, and 100-year return levels, respectively.} \label{table1}
\begin{center}
\begin{tabular}{ c l l l } 
 \hline
 Method & Delta & Proflik & Cond-sim \\ 
 \hline
 \hline
25-yr \, RL ECP  & 0.63 (46.9) & 0.80 (66.9) & 0.87 (111.1) \\ 
 \hline
50-yr \, RL ECP  & 0.74 (77.8) & 0.84 (93.4) & 0.87 (177.9) \\ 
 \hline
100-yr RL ECP  & 0.77 (107.8) & 0.82 (128.9) & 0.88 (295.4) \\ 
 \hline
\end{tabular}
\end{center}
\end{table}

\begin{figure}[H] 
\centering
\includegraphics[width=6in]{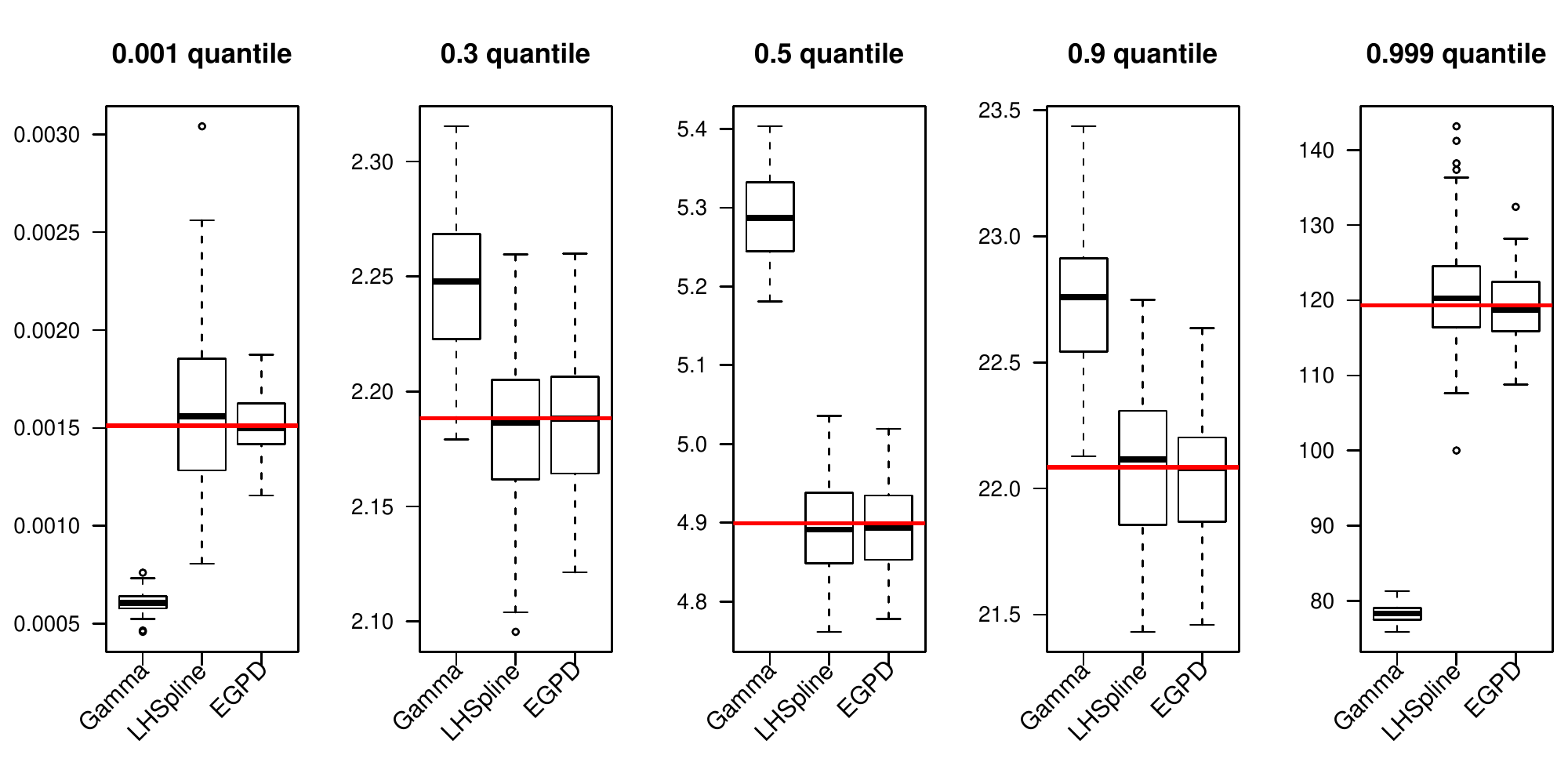}
\caption{Boxplots of quantile (from 0.001 to 0.999) estimates. The true values are represented by horizontal red lines.}
\label{fig:fig8}
\end{figure}

We also assess the estimation performance of LHSpline in the bulk distribution. Fig~\ref{fig:fig8} shows that the LHSpline performs much better than the gamma fit, a widely used distribution for modeling precipitation \citep{katz1977,wilks2011}, and performs nearly as well as the ``oracle'' (EGPD) approach.

\section{Applications} \label{sec4}

\subsection{Motivation: Hurricane Harvey Extreme Rainfall} \label{sec4.1}

\begin{figure}[H] 
\centering
\includegraphics[width=6in]{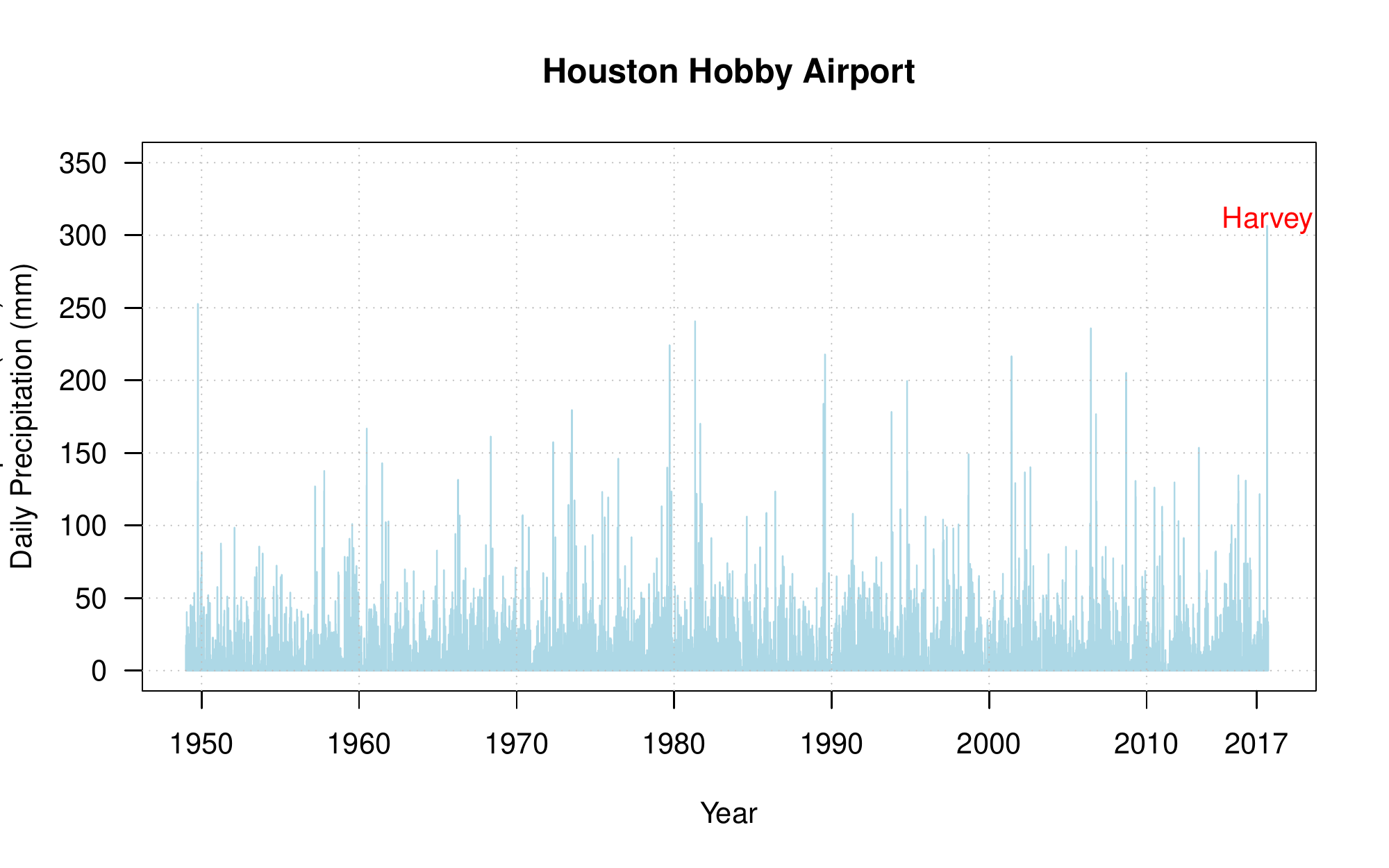}
\caption{The time series plot of Houston Hobby Airport daily precipitation from 1949 to 2017. Three out of the five largest daily precipitation values were observed during 26th to 28th August, 2017.}
\label{fig:fig9}
\end{figure}

Hurricane Harvey brought unprecedented amounts of rainfall to the Houston metropolitan area between the 25th and the 31st of August 2017, resulting in catastrophic damages to personal property and infrastructure. In the wake of such an extreme event, there is  interest in understanding its rarity and how human-induced
climate change might alter the chances of observing such an event \citep{risser2017}. In this section, we apply the LHSpline to the daily precipitation measurements for Houston Hobby airport (see Fig.~\ref{fig:fig1} for the histogram and Fig.~\ref{fig:fig9} for the time series) from the Global Historical  Climatology Network (GHCN) \citep{GHCN}.

\subsection{Houston Hobby Rainfall Data Analysis} \label{sec4.2}

\begin{figure}[H] 
\centering
\includegraphics[width=6in]{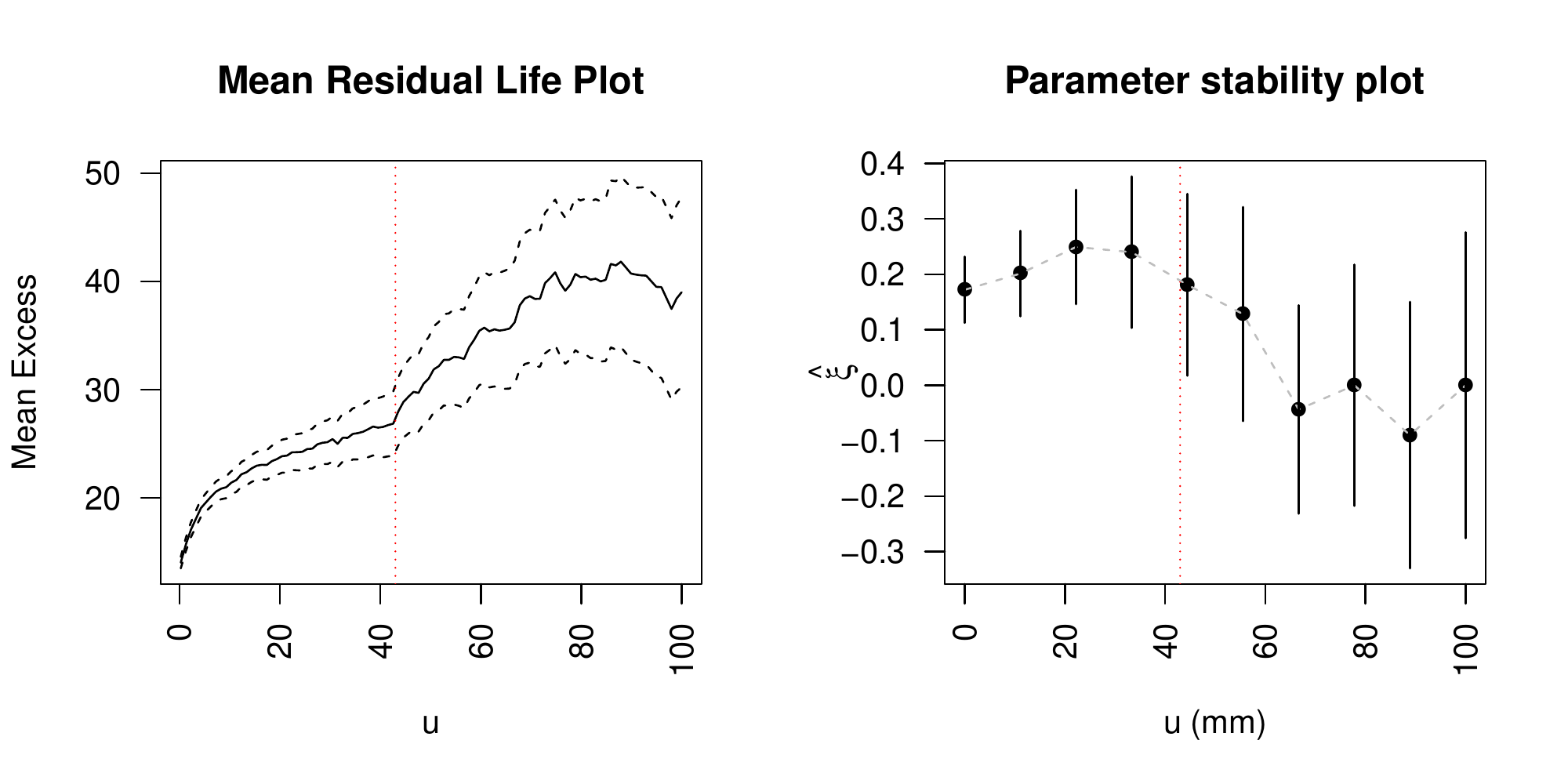}
\caption{\textbf{Left}: The mean residual life plot of the Hobby rainfall data. One should choose the smallest threshold, $u_{0}$, such that the mean excess, as a function of $u$, behaves linearly. \textbf{Right}: The parameter stability plot for the shape parameter of GPD of the Hobby rainfall. One should choose the $u_{0}$ such that the shape parameter estimates stabilize for $u > u_{0}$. The red vertical dotted line in each panel is the chosen threshold in this study.}
\label{fig:fig10}
\end{figure}

We fit a LHSpline with $150$ equally spaced bins to the full range of non-zero precipitation for Hobby Airport ($\sim 27.6\%$ of all the observations) prior to this event (from Jan. 1949 to Dec. 2016). To facilitate a comparison with the POT method we also fit a GPD to excesses over a high threshold (chosen as $u_{0} = 43 \text{ mm}$ $\sim .93$ empirical quantile of nonzero daily precipitation, see diagnostic plot in Fig.~\ref{fig:fig10}).\\  

\begin{figure}[H] 
\centering
\includegraphics[width=6.5in]{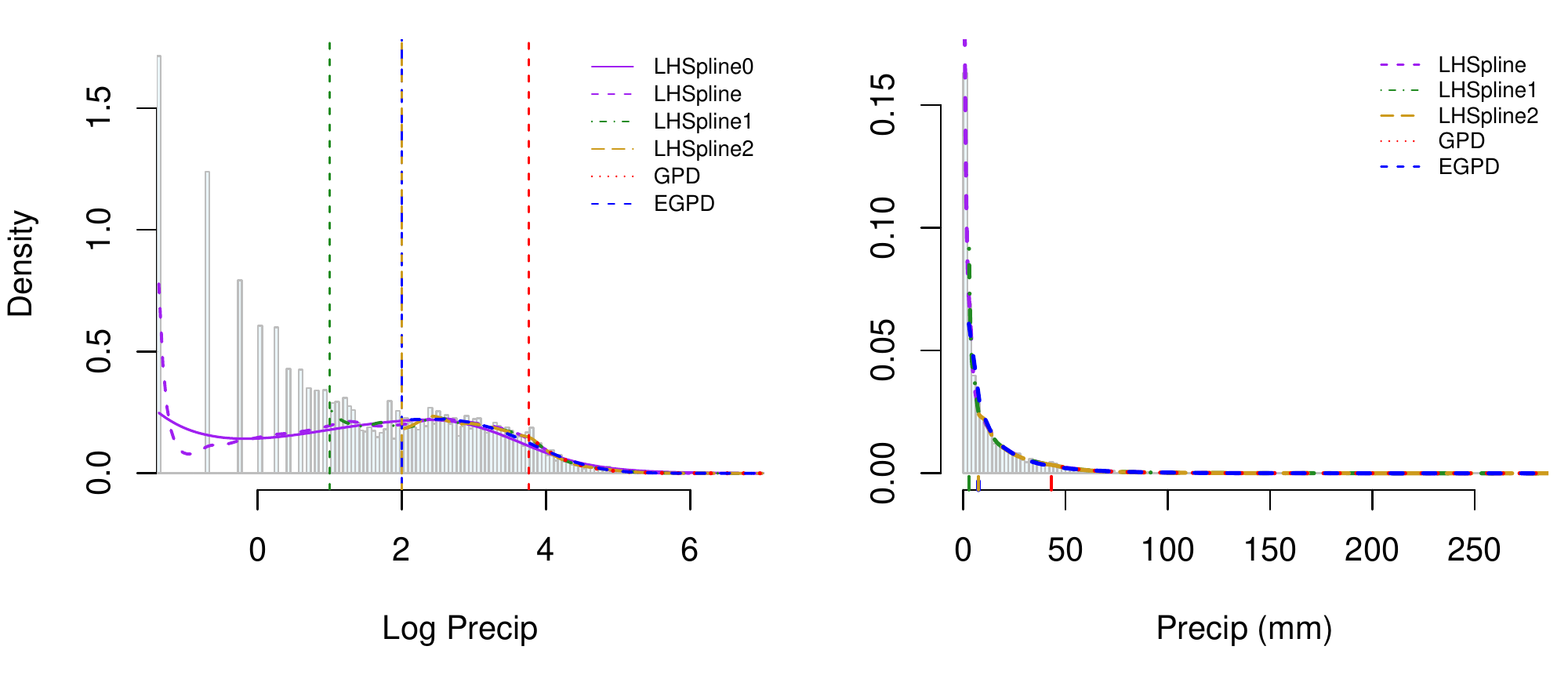}
\caption{The density estimate for log scale \textbf{(left)} and original scale \textbf{(right)} of the Hobby airport daily precipitation. The purple dashed (solid) lines are the estimates obtained by LHSpline with (without) $\lambda$ adjustment, green dot-dash by LHSpline with lower bound 1 ($\exp(1)$), golden longdash by LHSpline with lower bound 2 ($\exp(2)$), red dotted by GPD, and blue dotted by EGPD The red vertical lines indicate the threshold in the log scale (left) and the original scale (right). The ``discreteness'' of the histogram in the log scale is due to the precipitation measurement precision.}
\label{fig:fig11}
\end{figure}

An important practical issue when applying the LHSpline is that the discretization of rainfall measurement ($1/100^{\text{th}}$ of an inch) introduces an artifact in the finely binned histogram in the log scale (see Fig.~\ref{fig:fig11}, left panel) and hence the cross validation choice for $\lambda$ is affected by this discretization. In practice, all the precipitation measurements have precision limits and thus it is important to take this fact into account especially when the measurements are small (i.e., near zero). Here we treat the data being left-censored by truncating the values below two different values ($\exp(1)$ and $\exp(2)$) to alleviate this ``discreteness'' effect.

\begin{figure}[H] 
\centering
\includegraphics[width=6.5in]{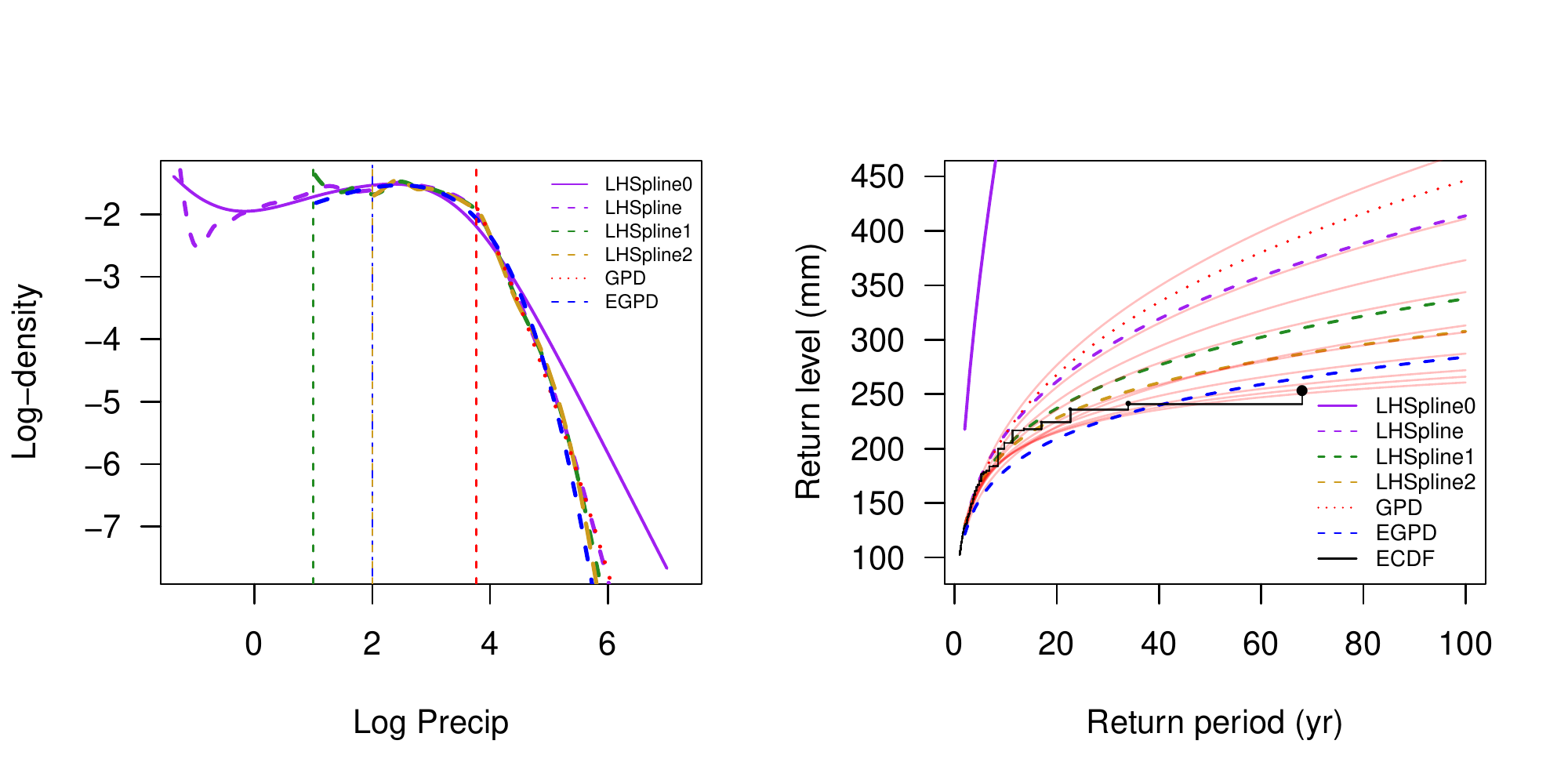}
\caption{The estimated log densities in the log scale \textbf{(left)} and return levels \textbf{(right)}. Light red lines are GPD estimates with different thresholds ranging from $.75$ to $.99$ empirical quantile. The black step function is the empirical return levels.}
\label{fig:fig12}
\end{figure}

As has been demonstrated in Fig.~\ref{fig:fig3} and Fig.~\ref{fig:fig4} in Sec.~\ref{sec3}, a good visual agreement in log-density (density) does not necessarily imply that it actually provides a good return level estimate. Fig.~\ref{fig:fig12} shows the estimated log-log plot and return levels ranging from 2 years to 100 years (under the assumption of temporal stationarity). The LHSpline gives somewhat lower estimates than that of the GPD estimates. A quantile-quantile plot in Fig.~\ref{fig:fig13} (left panel) indicates that there might be an issue of overestimating extreme high quantiles in the GPD fit. That is, the GPD estimate of the 68-year rainfall is 395.47 (90\% CI (300.88, 569.67)) mm whereas the ($\lambda$-adjusted) LHSpline estimates (with lower bound 0.254, $\exp(1)$, and $\exp(2$)) are 371.04 (302.39, 567.62) mm, 310.64 (261.62, 491.80) mm, and 287.06 (246.86, 457.05) mm, respectively, which are somewhat closer to the maximum value (252.73 mm) during the 1949 $\sim$ 2016 period. However, one should be aware that the sample maximum can be quite variable and hence might not reflect the true magnitude of the ``68-year rainfall''. Fig.~\ref{fig:fig13} (right panel) also indicates that the LHSpline might provide a better fit than the EGPD to the Hobby Airport data.    

\textbf{\begin{figure}[H] 
\centering
\includegraphics[width=6in]{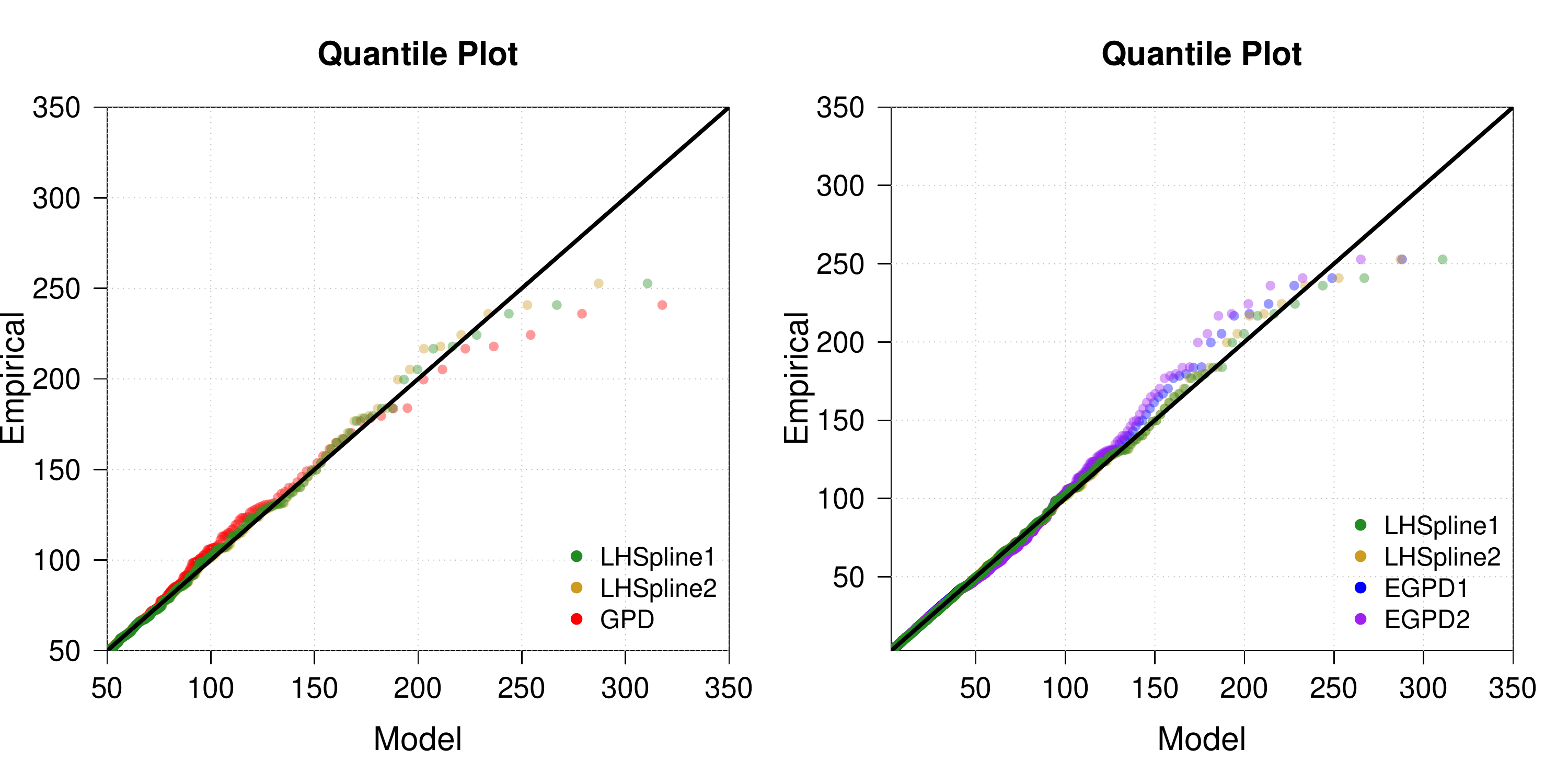}
\caption{The quantile-quantile plot of the Hobby daily precipitation data. \textbf{Left:} A comparison between LHSplines and GPD for the upper 5\% ($\ge 50 \text{ mm}$) of rainy data). \textbf{Right:} A comparison between LHSplines and EGPDs for the ``full range'' (left truncated at exp(1).}
\label{fig:fig13}
\end{figure}}

Lastly, we would like to investigate the question: ``\textit{How unusual was the event of 306.58 mm (26th August, 2017) in daily total precipitation at Hobby Airport}?'' To simplify the matter we make a stationarity assumption, that is, the distribution of daily precipitation does not change during the time period 1949 $\sim$ 2017. The estimates for the return period of this event are given in the following table:

\begin{table}[H]
\caption{The estimated return period of the 26th August, 2017 daily total precipitation observed at Hobby Airport using POT, LH-Spline, and EGPD.}\label{table2}
\begin{center}
\begin{tabular}{ c c c c c c c} 
 \hline
 Method & POT & LHSpline & LHSpline1 & LHSpline2& EGPD1 & EGPD2 \\ 
 \hline
 \hline
 Estimate (years) & 30.5 & 34.8 & 64.0 & 98.0 & 92.5 & 154.5 \\
90\% CI Lower limit & 17.0 & 15.2 & 19.1& 22.0& 71.5& 117.6\\
90\% CI Upper limit & 73.3 & 73.2 & 172.4& 345.7& 457.6& 515.6\\
 \hline
\end{tabular}
\end{center}
\end{table}

The much shorter return period estimate obtained from the POT method is due to the overestimation of the upper tail (see Fig.~\ref{fig:fig13}) whereas the somewhat longer return periods estimated by LHSplines might be more aligned with what people would expect. Although again one should be aware that there exists, among other issues, a large estimation uncertainty with respect to large return periods (see Table.~\ref{table2}).

\section{Discussion} \label{sec5}

This work presents a new statistical method, the LHspline, for estimating extreme quantiles of heavy-tailed distributions. In contrast with some widely used EVT based methods that require extracting ``extreme'' observations to fit a corresponding asymptotically justified distribution, the LHSpline makes use of the \textit{full range} of the observations for the fitting. By combining data transform and spline smoothing, the LHSpline estimation effectively achieved the desirable tail structure that is consistent with EVT and a flexible bulk distribution in the context of rainfall modeling. We demonstrate through simulation that this method
performs comparable to the POT method for return level estimation
with an additional benefit that it jointly models the bulk and the tail of a distribution.\\

However, by construction, the LHSpline method is only applicable for heavy-tailed distributions, which excludes many important environmental processes, for example, air surface temperature which may have a bounded tail \citep{gilleland2006}. In terms of implementation, the LHSpline requires some additional tuning, for example, the number of bins for the histogram should grow with the sample size. Limited experiments (not reported here) suggest the estimate is not sensitive to the number of bins once the binning is chosen fine enough. Another tuning issue is to decide how far one should extend beyond the range of the observations and how to adjust the smoothing parameter $\lambda$ to remove the boundary effect. Here we suggest extending the bin range and reducing the smoothing parameter. This choice is in place of picking the threshold in the POT approach and we believe it is less sensitive in terms of density estimation.\\

The theoretical properties of our method are largely unexplored. Much of the theoretical results developed in nonparametric density estimation concern the performance in terms of global indices such as $\mathbb{E}\left[\int_{x \in \mathbb{R}} |f(x) - \hat{f}(x)|\, dx\right]$ or $\mathbb{E}\left[\sup_{x \in  \mathbb{R}} |f(x) - \hat{f}(x)|\right]$ and are largely confined in a bounded region (i.e. $x \in [a,b]$, a and b finite). It is not clear how these results can inform us about the estimation performance for extreme quantiles on a potentially unbounded domain.\\

Applying LHSpline to many observational records or the grid cells of high resolution and/or ensemble climate model experiments \citep[e.g.][]{NARCCAP,wang2015} will result in summaries of the distribution that are well suited for further data mining and analysis. The log-density form is particularly convenient for dimension reduction because linear (additive) projection methods such as principle component analysis make sense in the log space.
%
%
%
In general we believe the LHSpline will be a useful tool as climate informatics tackles complex problems of quantifying climate extremes. 

\section*{Acknowledgements} \label{acknowledgements}
This work was conducted as part of the \href{https://www.statmos.washington.edu}{Research Network for Statistical Methods for Atmospheric and Oceanic Sciences} (STATMOS), supported by the National Science Foundation (NSF) Grants \#s 1106862, 1106974, and 1107046. This work was also supported by the NSF Grant \# 1638521 to the Statistical and Applied Mathematical Sciences Institute(SAMSI), the NSF Grant \# 1406536, and by the National Center for Atmospheric Research (NCAR). NCAR is sponsored by the NSF and managed by the University Corporation for Atmospheric Research. The authors would like to thank the anonymous reviewers for their valuable comments and suggestions to improve the quality of the paper.

\singlespacing
\bibliography{Loghistosp.bib}

\end{document}